\documentclass[10pt]{article}
\usepackage{bbm}
\usepackage[final]{graphics}
\usepackage{amsmath}
\usepackage{amsfonts,amsbsy}
\usepackage{amssymb}

\def\empile#1\over#2{\mathrel{\mathop{\kern 0pt#1}\limits_{#2}}}

\newcommand{\slv}{\raise.15ex\hbox{$/$}\kern-.53em\hbox{$v$}}
\newcommand{\slF}{\raise.15ex\hbox{$/$}\kern-.53em\hbox{$F$}}
\newcommand{\slL}{\raise.15ex\hbox{$/$}\kern-.53em\hbox{$L$}}
\newcommand{\slP}{\raise.15ex\hbox{$/$}\kern-.53em\hbox{$P$}}
\newcommand{\slp}{\raise.15ex\hbox{$/$}\kern-.53em\hbox{$p$}}
\newcommand{\slq}{\raise.15ex\hbox{$/$}\kern-.53em\hbox{$q$}}
\newcommand{\slR}{\raise.15ex\hbox{$/$}\kern-.53em\hbox{$R$}}
\newcommand{\slQ}{\raise.15ex\hbox{$/$}\kern-.53em\hbox{$Q$}}
\newcommand{\slK}{\raise.15ex\hbox{$/$}\kern-.53em\hbox{$K$}}
\newcommand{\slk}{\raise.15ex\hbox{$/$}\kern-.53em\hbox{$k$}}
\newcommand{\slD}{\raise.15ex\hbox{$/$}\kern-.53em\hbox{$D$}}
\newcommand{\slC}{\raise.15ex\hbox{$/$}\kern-.53em\hbox{$C$}}
\newcommand{\slA}{\raise.15ex\hbox{$/$}\kern-.53em\hbox{$A$}}
\newcommand{\slSigma}{\raise.15ex\hbox{$/$}\kern-.53em\hbox{$\Sigma$}}
\newcommand{\slpartial}{\raise.15ex\hbox{$/$}\kern-.53em\hbox{$\partial$}}
\newcommand{\slcalP}{\raise.15ex\hbox{$/$}\kern-.63em\hbox{$\cal P$}}

\def\p{{\boldsymbol p}}
\def\q{{\boldsymbol q}}

\def\x{{\boldsymbol x}}
\def\y{{\boldsymbol y}}

\catcode`\@=11

\newcount\@tempcntc
\def\@citex[#1]#2{\if@filesw\immediate\write\@auxout{\string\citation{#2}}\fi
  \@tempcnta\z@\@tempcntb\m@ne\def\@citea{}\@cite{%
        \@for\@citeb:=#2\do%
    {\@ifundefined{b@\@citeb}%
        {\@citeo\@tempcntb\m@ne\@citea%
                \def\@citea{,\penalty\@m\ }{\bf ?}\@warning%
                {Citation `\@citeb' on page \thepage \space undefined}}%
        {\setbox\z@\hbox{\global\@tempcntc0\csname b@\@citeb\endcsname\relax}
     \ifnum\@tempcntc=\z@ \@citeo\@tempcntb\m@ne%
       \@citea\def\@citea{,\penalty\@m}%
       \hbox{\csname b@\@citeb\endcsname}%
     \else%
      \advance\@tempcntb\@ne%
      \ifnum\@tempcntb=\@tempcntc%
      \else\advance\@tempcntb\m@ne\@citeo%
      \@tempcnta\@tempcntc\@tempcntb\@tempcntc\fi\fi}}\@citeo}{#1}}%

\def\@citeo{\ifnum\@tempcnta>\@tempcntb\else\@citea
  \def\@citea{,\penalty\@m}%
  \ifnum\@tempcnta=\@tempcntb\the\@tempcnta\else
   {\advance\@tempcnta\@ne\ifnum\@tempcnta=\@tempcntb \else
\def\@citea{--}\fi
    \advance\@tempcnta\m@ne\the\@tempcnta\@citea\the\@tempcntb}\fi\fi}

\catcode`\@=12

\begin{document}

\title{\bf Particle production in field theories\\ 
coupled to strong external sources\\
II. Generating functions}
\author{Fran\c cois Gelis$^{(1)}$, Raju Venugopalan$^{(2)}$}
\maketitle
\begin{center}
\begin{enumerate}
\item Service de Physique Th\'eorique (URA 2306 du CNRS)\\
  CEA/DSM/Saclay, B\^at. 774\\
  91191, Gif-sur-Yvette Cedex, France
\item Department of Physics, Bldg. 510 A,\\
Brookhaven National Laboratory,\\
  Upton, NY-11973, USA
\end{enumerate}
\end{center}

\begin{abstract}
We discuss a method for computing the generating function for the
multiplicity distribution in field theories with strong time dependent
external sources. At leading order, the computation of the generating
function reduces to finding a pair of solutions of the classical
equations of motion, with non-standard temporal boundary conditions.
\end{abstract}

\section{Introduction}

Ever since the advent of high energy colliders, the study of
multi-particle final states in hadronic collisions has offered the
possibility of probing the structure of QCD at a deep level.  Much
attention has been focused on the nature of multi-particle production
in jets -- for a nice review, see Ref.~\cite{DremiG1,WolfDK1}. The
problem is however very general. Theoretical developments in the last
couple of decades suggest that the bulk of multi-particle production
at the highest collider energies may be controlled by semi-hard
scales. This suggests the possibility that weak coupling computations
of final states can be compared to experiments. First steps in
applying this program to understanding high energy nuclear collisions
at the Relativistic Heavy Collider (RHIC) are very promising
\cite{BlaizG1}. We expect these considerations to be even more
relevant for next generation experiments at RHIC and at the Large
Hadron Collider (LHC).

Motivated by these considerations, in a previous paper \cite{GelisV2},
we developed a general formalism to study the properties of particle
production in a field theory coupled to a strong external
time-dependent source. We focused there, for simplicity, on the case
of a $\phi^3$ scalar field theory. We believe however that most of our
results are of general validity and can be extended to a gauge theory
like QCD.

The relevance of such a field theory coupled to strong sources to the
description of multi-particle final states in high energy hadronic
interactions comes from the possibility of separating, in a high energy
hadron wave-function, partons with a large longitudinal momentum
fraction $x$ from low $x$ partons. This is because large $x$ partons
are time-dilated and evolve very little over the interaction time with
the other hadron. They are therefore ``frozen" color sources for
partons at smaller $x$.  The latter, in contrast, are dynamical fields
over the time scales of the interaction process. The
Balitsky-Fadin-Kuraev-Lipatov (BFKL) evolution equation
\cite{BalitL1,KuraeLF1} predicts that parton densities grow very
rapidly with decreasing $x$. Because this rapid growth leads to large
phase-space densities of partons in the hadron wave-function, the
dynamics of the field that describes the small $x$ partons is to a
good approximation classical. This separation of the degrees of
freedom of a high energy hadron into static sources and dynamical
fields that behave almost classically is the essence of the
McLerran-Venugopalan model \cite{McLerV1,McLerV2,McLerV3}.  The
corresponding effective theory is the theory we mimicked in
\cite{GelisV2}.

In a high energy hadron, the effective theory that results from this
separation of the degrees of freedom is completely specified by the
knowledge of the distribution of hard sources. Because physical
results should not depend on the arbitrary scale which is used to
separate the degrees of freedom, the distribution of sources must obey
a renormalization group equation, known as the JIMWLK equation
\cite{JalilKMW1,JalilKLW1,JalilKLW2,JalilKLW3,JalilKLW4,IancuLM1,IancuLM2,FerreILM1}.
In the limit of a large number of colors and of a large nucleus, it
takes a much simpler form, the Balitsky-Kovchegov equation
\cite{Balit1,Kovch3}. This framework is referred to as the Color Glass
Condensate (CGC) \cite{McLer1,IancuLM3,IancuV1}.

In this framework, to compute an observable in a high energy hadronic
collision, one first finds the appropriate distribution of color
sources and subsequently computes the observable of interest for that
particular distribution of color sources. Finally, the observable has
to be averaged over all possible configurations of distributed
sources. The former involves solving the evolution equation for the
distribution of sources. As in \cite{GelisV2} we will assume here that
the distribution of sources is known and focus on the second part of
the calculation. We will also assume that factorization holds and that
any back reaction from the produced fields on the sources can be
neglected. Finally, we will assume that the sources (because of the
rapid growth of parton densities with decreasing $x$) are strong
sources. This growth is tamed only when the self-interactions of the
partons become important, a regime which is called ``saturation''
\cite{GriboLR1,MuellQ1,BlaizM1,Muell4}.  In this saturation regime,
which is natural for a non-Abelian gauge theory at sufficiently high
energies, the color sources are of the order of the inverse of the
coupling constant. They are therefore strong sources in a weakly
coupled theory.

A crucial feature of the strong source saturation regime is that the
calculation of any observable in the background provided by such
sources is non perturbative in the specific sense that one needs to
sum an infinite set of Feynman diagrams, even at leading
order\footnote{Note that in collisions involving at least one
``dilute'' projectile (proton-nucleus collisions at moderate energies,
or even nucleus-nucleus collisions near the fragmentation region of
one of the nuclei), the calculation of observables can be carried out
analytically
\cite{KovchM3,DumitM1,DumitJ1,DumitJ2,GelisJ1,GelisJ2,GelisJ3,GelisV1,BlaizGV1,BlaizGV2}.}.

One therefore needs to develop techniques in order to sum the relevant
diagrams. In \cite{GelisV2}, we showed how the calculation of the
average number of produced particles can be remapped, at leading order
in the coupling constant, into the problem of solving the classical
equation of motion (EOM) for the field with retarded boundary conditions. We
also showed that the average multiplicity could be computed at
next-to-leading order (NLO) by solving in addition the equation of
motion for a small fluctuation of the field on top of the classical
solution. Again, remarkably, this is an initial value problem with
purely retarded boundary conditions.  The retarded solutions of these
two equations of motion also allows one in principle to obtain higher
moments of the distribution of multiplicities. One such quantity is
the variance of the distribution of particles.

This framework was already employed previously to compute gluon
production at leading order
\cite{KrasnV4,KrasnV1,KrasnV2,KrasnNV1,KrasnNV2,Lappi1} and in the
calculation of quark production \cite{GelisKL1,GelisKL2}. The latter
computation, leading order in quark production, is similar to gluon
pair production, which contributes to the gluon multiplicity at NLO. A
recent numerical study of plasma instabilities in heavy ion collisions
\cite{RomatV1,RomatV3} is very similar in spirit to the computation of
the small fluctuation fields necessary to determine NLO contributions
to the average multiplicity. One should emphasize a crucial aspect of
these results: the equations of motion are solved with retarded
boundary conditions. {\sl This allows one to develop straightforward
numerical algorithms for finding these solutions even at
next-to-leading order}.

In \cite{GelisV2}, we introduced a generating function for the
distributions of multiplicities. We only used it there as an
intermediate device for finding expressions for the moments of the
distribution of multiplicities. In fact, determining it would allow one
to obtain the distribution of probabilities for $n$-particle final
states rather than the average multiplicity alone. We shall discuss
here the problem of computing the generating function for particle
production. Expressing the generating function as a sum of
vacuum-vacuum diagrams in the Schwinger-Keldysh formalism
\cite{Schwi1,Keldy1}, we show that the derivative of the generating
function is given by a sum of diagrams that are very similar to those
involved in the calculation of the average multiplicity. Further
closely examining the set of relevant diagrams at leading order, we
show that their sum can be expressed in terms of two solutions of the
classical equation of motion. However, unlike the average
multiplicity, the boundary conditions obeyed by these solutions are
not retarded and instead, one has boundary conditions both at the
initial and at the final time. This makes the numerical solution of
this problem much more difficult. We will discuss briefly one possible
strategy for its numerical solution.

The paper is organized as follows. In section \ref{sec:reminder}, we
remind the reader of results derived in \cite{GelisV2} that are
important later in our discussion. In section \ref{sec:all-orders}, we
define the generating function and establish a formula for its
derivative. This formula is valid to all orders in the coupling
constant.  In section \ref{sec:LO}, we restrict ourselves to leading
order; at this order, the derivative of the generating function can be
expressed in terms of solutions of the classical equation of
motion. We derive the boundary conditions for these solutions. These
are shown to be very simple constraints on the fields at times $\pm
\infty$. We briefly discuss a possible strategy to compute these
solutions numerically. In section \ref{sec:variance}, we use the
previous results in order to derive an explicit formula for the
variance of the number of produced particles, in terms of solutions of
the equation of motion for small field fluctuations on top of the
classical field. Finally, in section \ref{sec:E-GF}, we construct a
generating function for the distribution of produced energy as opposed
to the number of particles. Such a generating function is likely more
relevant to gauge theories that have infrared problems. We show that
it can be obtained at leading order (in a very similar fashion to the
generating function for the number) from solutions of the classical
EOM.

\section{Reminder of some results of \cite{GelisV2}}
\label{sec:reminder}

In this section, we shall provide a brief synopsis of the results in
\cite{GelisV2} on computing moments of the multiplicity distribution
in a field theory with time dependent external sources.
\subsection{Model field theory}
\label{sec:model}
The Lagrangian density in our toy model is
\begin{equation}
{\cal L}\equiv\frac{1}{2}\partial_\mu\phi\,\partial^\mu\phi 
-\frac{1}{2}m^2\phi^2-\frac{g}{3!}\phi^3 +j\phi\; .
\label{eq:lagrangian}
\end{equation}
We assume the source $j(x)$ to be time-dependent and of strength
$1/g$. By that we mean that the dimensionless number $\int d^4 x\, g\,j(x)\sim O(1)$ 
when we do the power counting for a diagram~\footnote{Note that a weak coupling approach in 
this model is only valid for $m^2 > 0$. Else, the theory has no minimum about which a stable 
weak coupling expansion can be performed.}. 

\subsection{Direct calculation of $P_n$}
When the current $j(x)$ coupled to the fields is time-dependent,
 non-zero transition amplitudes between the vacuum and
populated states are allowed.  The probability $P_n$ for the
production of $n$ particles is 
\begin{equation}
P_n=\frac{1}{n!}\int\prod_{i=1}^n\frac{d^3\p_i}{(2\pi)^3 2E_i}
\;
\Big|
\big<\p_1\cdots\p_n\,{}_{\rm out}\big|0_{\rm in}\big>
\Big|^2\; .
\end{equation}
It is well known that the transition amplitude
$\big<\p_1\cdots\p_n\,{}_{\rm out}\big|0_{\rm in}\big>$ is the sum of
all the Feynman diagrams with only sources in the initial state and
$n$ particles in the final state. Note that these Feynman diagrams are
built from {\bf time-ordered} propagators. In a field theory coupled
to a time-dependent source, one should also include the disconnected
vacuum-vacuum diagrams in computing the probabilities.  (This is in
sharp contrast to conventional field theories without
external sources where such diagrams can indeed be ignored because
they add up to a pure phase in the amplitude and unity in the
probabilities.)

\begin{figure}
\begin{center}
\resizebox*{!}{5cm}{\includegraphics{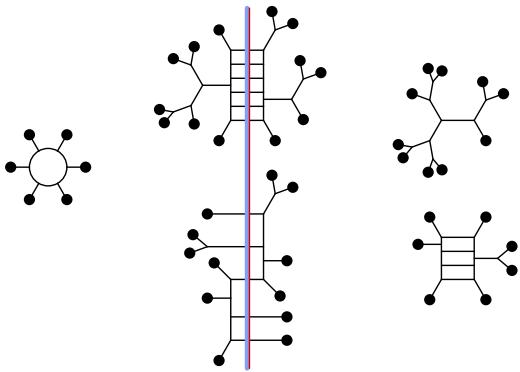}}
\end{center}
\caption{\label{fig:P11}A typical contribution to $P_{11}$, the
probability to produce a final state with 11 particles. The vertical
cut delimits the amplitude and its complex conjugate.}
\end{figure}

The necessity of keeping track of the vacuum-vacuum diagrams is what
makes the direct calculation of any of the probabilities $P_n$
difficult. In order to illustrate the difficulty of the task, we
display in figure \ref{fig:P11} a typical contribution to $P_{11}$,
the probability of producing 11 particles in the final state.
The number of disconnected vacuum-vacuum graphs is arbitrary, although
their sum is the exponential of all the connected ones:
\begin{equation}
i\sum_{\rm all}V = \exp \left[i\sum_{\rm conn}V\right]\; .
\end{equation}
In a field theory in the vacuum, $\sum_{\rm conn}V$ is a real number,
and $\exp\big( i\sum_{\rm conn}V\big)$ and the similar factor from the
complex conjugate amplitude cancel each other. However, this is not
the case in the presence of a time-dependent source $j(x)$. One should
also note that it not possible to perform a perturbative expansion of
the probability $P_n$ when the strength of the source is
$1/g$. Indeed, the leading order contribution in this approach
corresponds to keeping only the {\bf tree} disconnected factors, but
because each of these factors is of order $1/g^2$, any truncation is
meaningless.

\subsection{Calculation of the moments}
In \cite{GelisV2}, we obtained a very compact formula for the
probability $P_n$, that reads
\begin{equation}
P_n
=
\frac{1}{n!}
{\cal D}^n[j_+,j_-]\;
\left.
e^{iV[j_+]}\;e^{-iV^*[j_-]}
\right|_{j_+=j_-=j}\; .
\label{eq:Pn-compact}
\end{equation}
In this formula, $iV[j_+]$ is the sum of all the connected
time-ordered vacuum-vacuum diagrams, constructed with the source
$j_+$, and $-iV^*[j_-]$ is its complex conjugate, constructed with the
source $j_-$. $D[j_+,j_-]$ is an operator that acts on the sources
$j_+,j_-$, defined as
\begin{equation}
{\cal D}[j_+,j_-]\equiv
\frac{1}{Z}\int d^4x\, d^4y\; G_{+-}^0(x,y)\;
(\square_x+m^2)(\square_y+m^2)\;
\frac{\delta}{\delta j_+(x)}\,\frac{\delta}{\delta j_-(y)}\; ,
\end{equation}
where $Z$ is the wave function renormalization constant and
$G_{+-}^0(x,y)$ is the $+-$ propagator of the
Schwinger-Keldysh \cite{Schwi1,Keldy1} formalism\footnote{See 
appendix A of \cite{GelisV2} for a brief reminder on the
Schwinger-Keldysh formalism.}
\begin{equation}
G_{+-}^0(x,y)\equiv\int\frac{d^3\p}{(2\pi)^32E_\p}\;e^{ip\cdot(x-y)}\; .
\end{equation}

Eq.~(\ref{eq:Pn-compact}), although very formal and of no use for
calculating $P_n$ itself, enables one to obtain very simply formulas
for the moments.  From these, one gets easily the average multiplicity
\begin{equation}
\big<n\big>\equiv\sum_{n=0}^{\infty}nP_n=
{\cal D}[j_+,j_-]\;
\left.
e^{{\cal D}[j_+,j_-]}\;
e^{iV[j_+]}\;e^{-iV^*[j_-]}
\right|_{j_+=j_-=j}\; .
\label{eq:nbar-1}
\end{equation}
A crucial observation in \cite{GelisV2} was that the product
\begin{equation}
e^{{\cal D}[j_+,j_-]}\;
e^{iV[j_+]}\;e^{-iV^*[j_-]}\equiv e^{i{\cal V}_{_{SK}}[j_+,j_-]}\, ,
\end{equation}
namely, is the sum of the vacuum-vacuum diagrams of the Schwinger-Keldysh
formalism, with the source $j_+$ on the upper branch of the time
contour, and $j_-$ on the lower branch.

From eq.~(\ref{eq:nbar-1}), one obtains, 
for $\big<n\big>$
\begin{equation}
\big<n\big>
=
\int d^4x d^4y\;
ZG_{+-}^0(x,y)
\;\left[
\Gamma^{(+)}(x)\Gamma^{(-)}(y)+\Gamma^{(+-)}(x,y)
\right]_{j_+=j_-=j}\; ,
\label{eq:n3}
\end{equation}
where $\Gamma^{(\pm)}$ and $\Gamma^{(+-)}$ are the 1- and 2-point
amputated Green's functions in the Schwinger-Keldysh formalism:
\begin{eqnarray}
&&
\Gamma^{(\pm)}(x)\equiv \frac{\square_x+m^2}{Z}\;
\frac{\delta i{\cal V}_{_{SK}}[j_+,j_-]}{\delta j_\pm(x)}\; ,
\nonumber\\
&&
\Gamma^{(+-)}(x,y)
\equiv
\frac{\square_x+m^2}{Z}\;\frac{\square_y+m^2}{Z}\;
\frac{\delta^2 i{\cal V}_{_{SK}}[j_+,j_-]}{\delta j_+(x)\delta j_-(y)}\; .
\end{eqnarray}
Diagrammatically, $\big<n\big>$ can be represented as
\setbox1\hbox to 4cm{\hfil\resizebox*{4cm}{!}{\includegraphics{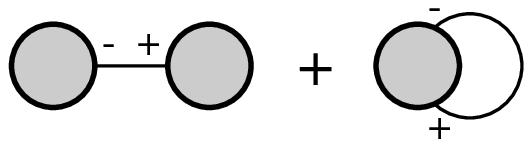}}}
\begin{equation}
\big<n\big>=\quad\raise -4.5mm\box1\quad .
\label{eq:nbar1}
\end{equation}
Note that, contrary to the individual probabilities $P_n$, the average
multiplicity is composed only of connected diagrams.

A major simplification occurs in this formula at leading order. {\sl
When $j_+=j_-=j$, the sum of the 1-point tree diagrams of the
Schwinger-Keldysh formalism is equal to the {\bf retarded} solution of
the classical equation of motion, with a vanishing boundary condition
at $x_0=-\infty$.} The problem of summing all the diagrams that
contribute to $\big<n\big>$ at leading order therefore maps into the
problem of finding the solution of a partial differential
equation. Moreover, because the boundary conditions for this equation
are retarded, the problem is straightforward to solve numerically.
Further, we also showed in \cite{GelisV2} that, at next-to-leading
order, the calculation of $\big<n\big>$ only requires one to solve the
equation of motion for small fluctuations of the field on top of the
classical solution again, with retarded boundary conditions. This classical 
solution was previously obtained when computing $\big<n\big>$ at
leading order.

The simplifications observed for the first moment, the average
multiplicity $\big<n\big>$, are quite generic. For the moment of order
$p$, one can derive a formula similar to eq.~(\ref{eq:nbar1}),
although these formulas become increasingly cumbersome as the order
$p$ increases. (The formula for $p=2$ was given explicitly in
\cite{GelisV2}, and we shall not reproduce it here.) Following the
same arguments as those for the average multiplicity, all moments can
in principle be calculated from the retarded solution of the classical
EOM plus retarded solutions of the small fluctuation EOM. Of course,
doing so in practice is prohibitively computing intensive with
increasing $p$.

\section{Generating function: general features}
\label{sec:all-orders}
\subsection{Definition}

The generating function is simply defined as\footnote{In \cite{GelisV2}, we defined it instead
as
\begin{equation}
F(x)\equiv \sum_{n=0}^{+\infty} P_n\, e^{nx}\; .
\end{equation}
This alternative definition was chosen because it simplified the calculation
of the moments of the distribution of multiplicities
\begin{equation}
\big<n^p\big>
\equiv
\sum_{n=0}^{+\infty} P_n\,n^p=F^{(p)}(0)\; ,
\end{equation}
to the $p$th derivative of the generating function. The correspondence
between the definition in this paper and the one used in \cite{GelisV2} is of
course $F(x)={\cal F}(e^x)$.}
\begin{equation}
{\cal F}(z)\equiv \sum_{n=0}^{+\infty}P_n\, z^n\; ,
\label{eq:cal-F-def}
\end{equation}
where the variable $z$ is possibly complex.  

In \cite{GelisV2}, the generating function was purely an intermediate
device to simplify the derivation of moments of the multiplicity
distribution. The generating function itself was never
computed. However, by calculating ${\cal F}(z)$, one would have access
to the complete sequence of probabilities $P_n$; assuming ${\cal
F}(z)$ is known, it is very easy to go back to the probabilities $P_n$
by an integration in the complex plane\footnote{From the definition in
eq.~(\ref{eq:cal-F-def}), the probabilities $P_n$ can also be
determined from the generating function by taking successive
derivatives at $z=0$~: $P_n={\cal F}^{(n)}(0)/n!$. However, the
interesting values of $n$ are located around the average value
$\big<n\big>$.  Because the typical multiplicity is very large in a
heavy ion collision, derivatives of very high order would have to be
calculated for this method to apply.  This is difficult to do
numerically with a good precision.}~:
\begin{equation}
P_n=\frac{1}{2\pi i}\oint_{\cal C} \,\frac{dz}{z^{n+1}} \, {\cal F}(z)\; ,
\end{equation}
where ${\cal C}$ is a closed path circling around the origin $z=0$. 

Choosing the contour ${\cal C}$ to be the unit circle, we can obtain
$P_n$ very effectively as a Fourier coefficient\footnote{In practice,
one should write the Fourier integral as the discrete Riemann sum
\begin{equation*}
P_n=\lim_{N\to+\infty}\frac{1}{2\pi N}\sum_{k=0}^{N-1}
e^{-2i\pi \frac{kn}{N}}\;{\cal F}(e^{2i\pi\frac{k}{N}})\; ,
\end{equation*}
and evaluate the sum by using the ``fast Fourier transform'' (FFT)
algorithm. Note that the maximal ``frequency'' $N$ one uses in the FFT
limits the largest multiplicity $n$ which is accessible.} of the
function ${\cal F}(e^{i\theta})$
\begin{equation}
P_n 
=\frac{1}{2\pi}\int_0^{2\pi} d\theta\; e^{-in\theta}\;{\cal F}(e^{i\theta})\; .
\end{equation}

\begin{figure}
\begin{center}
\resizebox*{8cm}{!}{\rotatebox{-90}{\includegraphics{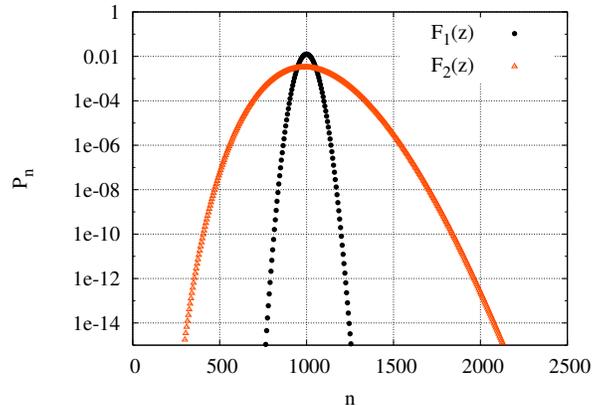}}}
\end{center}
\caption{\label{fig:Pn}Distribution of multiplicities for two toy models of
the generating function ${\cal F}(z)$. Solid circles~: ${\cal
F}_1(z)=\exp(\overline{n}(z-1))$. This generates a Poisson
distribution of average $\overline{n}$ -- we set
$\overline{n}=1000$ in the figure. Open triangles~: ${\cal
F}_2(z)=\exp(\overline{n}(\frac{z-1}{4} + \frac{z^{10}-1}{40} +
\frac{z^{20}-1}{80} + \frac{z^{30}-1}{120}))$. In this second model,
$25\%$ of the particles are produced individually, $25\%$ of the
particles come in bunches of 10, $25\%$ in bunches of 20 and the
remaining $25\%$ in bunches of 30. The average $\overline{n}$ is the
same as in the first model.}
\end{figure}

The use of this formula to go from the generating function $F(z)$ to
the distribution of probabilities $P_n$ is illustrated in figure
\ref{fig:Pn} with two toy models for ${\cal F}(z)$.
The two models have the same average number of produced particles, but
differ vastly in how these particles are correlated. As one can see,
the effect of the correlations in the latter case is to significantly
alter the width of the multiplicity distribution. It is precisely
because the average number $\big<n\big>$ alone cannot discriminate
between these very different production mechanisms that one wishes to
devise methods to obtain more detailed information about the
distribution of produced particles.


\subsection{${\cal F}(z)$ as a sum of vacuum-vacuum diagrams}
From eqs.~(\ref{eq:Pn-compact}) and ~(\ref{eq:cal-F-def}), we obtain simply,
\begin{equation}
{\cal F}(z)=
\left.
e^{z{\cal D}[j_+,j_-]}\;e^{iV[j_+]}\; e^{-iV^*[j_-]}
\right|_{j_+=j_-=j}\; .
\label{eq:genF-1}
\end{equation}
This expression has a diagrammatic interpretation as the sum of the
vacuum-vacuum diagrams of the Schwinger-Keldysh formalism, with each
off-diagonal propagator ($+-$ or $-+$) weighted by a factor $z$. This
is equivalent to modifying the off-diagonal Schwinger-Keldysh
propagators by 
\begin{eqnarray}
&&
G_{+-}^0\quad\longrightarrow\quad z\;G_{+-}^0\; ,
\nonumber\\
&&
G_{-+}^0\quad\longrightarrow\quad z\;G_{-+}^0\; .
\label{eq:modif-SK}
\end{eqnarray}
 We shall exploit this property later when we propose a method for
calculating the generating function. For the record, the expressions
for all the propagators involved in the diagrammatic expansion of the
generating function are
\begin{eqnarray}
&&
G^0_{++}(x,y)=\int\frac{d^3\p}{(2\pi)^3 2E_\p}
\left\{
\theta(x^0-y^0)e^{-ip\cdot(x-y)}
+
\theta(y^0-x^0)e^{ip\cdot(x-y)}
\right\}\; ,
\nonumber\\
&&
G^0_{--}(x,y)=\int\frac{d^3\p}{(2\pi)^3 2E_\p}
\left\{
\theta(x^0-y^0)e^{ip\cdot(x-y)}
+
\theta(y^0-x^0)e^{-ip\cdot(x-y)}
\right\}\; ,
\nonumber\\
&&
zG^0_{+-}(x,y)=z\int\frac{d^3\p}{(2\pi)^3 2E_\p}
\;e^{ip\cdot(x-y)}\; ,
\nonumber\\
&&
zG^0_{-+}(x,y)=z\int\frac{d^3\p}{(2\pi)^3 2E_\p}
\;e^{-ip\cdot(x-y)}\; ,
\label{eq:props}
\end{eqnarray}
where $E_\p\equiv\sqrt{\p^2+m^2}$ is the on-shell energy, and
implicitly $p_0=+E_\p$.

\subsection{All-orders result}
The sum of all the vacuum-vacuum diagrams involved in ${\cal F}(z)$ is
the exponential of the sum that contains only the connected ones. It can be 
written as 
\begin{equation}
e^{z{\cal D}[j_+,j_-]}\;
\;e^{iV[j_+]}\; e^{-iV^*[j_-]}
\equiv
e^{i{\cal V}_{_{SK}}[z|j_+,j_-]}\; ,
\label{eq:F-gen-1}
\end{equation}
where ${\cal V}_{_{SK}}[z|j_+,j_-]$ is the generalization to $z\not=1$
of 
\begin{equation}
{\cal V}_{_{SK}}[j_+,j_-]\equiv {\cal V}_{_{SK}}[1|j_+,j_-]\; ,
\end{equation}
 encountered earlier in the calculation of $\big<n\big>$. The
generating function is simply obtained from this quantity by equating
$j_+$ and $j_-$~:
\begin{equation}
{\cal F}(z)=e^{i{\cal V}_{_{SK}}[z|j,j]}\; .
\end{equation}

Let us now consider the derivative of the generating function with
respect to its argument $z$. We can differentiate directly
eq.~(\ref{eq:genF-1}) with respect to $z$, which leads to
\begin{equation}
{\cal F}^\prime(z)
={\cal D}[j_+,j_-]\;\left.
e^{i{\cal V}_{_{SK}}[z|j_+,j_-]}
\right|_{j_+=j_-=j}\; .
\label{eq:genF-gen-der-2}
\end{equation}
As previously in the calculation of $\big<n\big>$, making explicit the action
of the operator ${\cal D}[j_+,j_-]$, one writes 
\begin{equation}
{\cal F}^\prime(z)
=
e^{i{\cal V}_{_{SK}}[z|j,j]}
\;
\int d^4x d^4y \;ZG^0_{+-}(x,y)
\Big[
\Gamma^{(+)}(z|x)\Gamma^{(-)}(z|y)
+\Gamma^{(+-)}(z|x,y)
\Big]\; ,
\end{equation}
where we have defined
\begin{eqnarray}
&&
\Gamma^{(\pm)}(z|x)\equiv \frac{\square_x+m^2}{Z}\;
\left.
\frac{\delta i{\cal V}_{_{SK}}[z|j_+,j_-]}{\delta j_\pm(x)}
\right|_{j_+=j_-=j}
\nonumber\\
&&
\Gamma^{(+-)}(z|x,y)\equiv \frac{\square_x+m^2}{Z}\;\frac{\square_y+m^2}{Z}\;
\left.
\frac{\delta^2 i{\cal V}_{_{SK}}[z|j_+,j_-]}
{\delta j_+(x)\,\delta j_-(y)}
\right|_{j_+=j_-=j}
\; .
\end{eqnarray}
Therefore, one can write the derivative as 
\begin{equation}
\frac{{\cal F}^\prime(z)}{{\cal F}(z)}
=
\int d^4x d^4y \;Z\,G^0_{+-}(x,y)
\Big[
\Gamma^{(+)}(z|x)\Gamma^{(-)}(z|y)
+\Gamma^{(+-)}(z|x,y)
\Big]\; .
\label{eq:deriv-gen-func}
\end{equation}

This relation is very similar in structure to the formula we 
derived previously in \cite{GelisV2} for the average multiplicity
$\big<n\big>$. In fact, it contains exactly the same
topologies\footnote{This similarity is not a coincidence. Indeed, 
\begin{equation*}
\big<n\big>=\frac{{\cal F}^\prime(1)}{{\cal F}(1)}\; .
\end{equation*}
}, \setbox1\hbox to
4cm{\hfil\resizebox*{4cm}{!}{\includegraphics{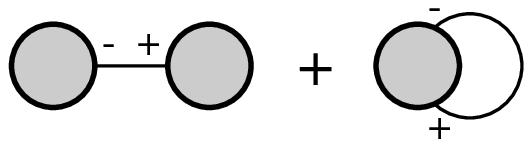}}}
\begin{equation}
\frac{{\cal F}^\prime(z)}{{\cal F}(z)}
=\quad\raise -4mm\box1\quad ,
\label{eq:derV1}
\end{equation}
except that the 1-point and 2-point Greens functions
$\Gamma^{(\pm)}(z|x)$ and $\Gamma^{(+-)}(z|x,y)$ must be evaluated
with modified -- $z$-dependent -- Schwinger-Keldysh rules. It is
precisely this similarity with $\big<n\big>$ which makes the formula
very attractive. As we shall see shortly, at leading order, ${{\cal
F}^\prime(z)}/{{\cal F}(z)}$ can be expressed in terms of solutions of
the classical equation of motion. If one succeeds in calculating it,  
${\cal F}(z)$ can be subsequently determined from the relation\footnote{If
${{\cal F}^\prime(z)}/{{\cal F}(z)}$ does not have poles, then we can
integrate along any path in the complex plane that goes from $1$ to
$z$. If on the contrary this quantity has a pole at some $z_0$, then
$\ln{\cal F}(z)$ itself has a logarithmic branch cut starting at
$z_0$. The result of the integration now depends on the number of
windings of the integration path around the pole $z_0$, an ambiguity
which amounts to choosing the Riemann sheet in which the point $z$
lies. Note however that this does not introduce any ambiguity in
${\cal F}(z)$ itself.}
\begin{equation}
\ln{\cal F}(z)=
\ln{\cal F}(1)
\;+\int_1^z
dz^\prime\;
\frac{{\cal F}^\prime(z^\prime)}{{\cal F}(z^\prime)}\; .
\label{eq:integration}
\end{equation}
 The integration constant
$\ln{\cal F}(1)$ can be determined trivially from the unitarity condition
\begin{equation}
{\cal F}(1)=\sum_{n=0}^\infty P_n=1\; .
\end{equation}
Thus,
\begin{equation}
{\cal F}(z)=
\exp\left\{
\int_1^z
dz^\prime\;
\frac{{\cal F}^\prime(z^\prime)}{{\cal F}(z^\prime)}\right\}\; .
\label{eq:integration1}
\end{equation}

\section{Generating function at leading order}
\label{sec:LO}
An urgent question at this point is whether there exists a practical
way to calculate the ratio ${{\cal F}^\prime(z)}/{{\cal F}(z)}$, given
the strong similarities between the formulas for ${{\cal
F}^\prime(z)}/{{\cal F}(z)}$ and for $\big<n\big>$. Because there is a
well defined algorithm to compute $\big<n\big>$, order by order, in
terms of solutions of classical equations of motion with retarded
boundary conditions, it might be hoped that a similar algorithmic
procedure exists for ${{\cal F}^\prime(z)}/{{\cal F}(z)}$.

In this section, we explore this issue at leading order. It is
important to stress at the outset that by ``leading order'' we refer
to the logarithm of the generating function, or of its derivative with
respect to the variable $z$. For these two quantities, the leading
order is the order $1/g^2$. The ``next-to-leading order'' would be the
order $g^0$, and so on. Discussing the order of the generating
function ${\cal F}(z)$ itself does not make sense because it is
evident that the exponentiation of the logarithms will mix up terms at
all orders. This mixing was previously encountered in the
probabilities $P_n$ themselves\footnote{See eq.~(39) of
\cite{GelisV2}, where all the numbers $b_r$, as well as the $a$ in the
exponential, are of order $1$ in the leading order approximation}.

\subsection{Expression for 
${{\cal F}^\prime(z)}/{{\cal F}(z)}$ at leading order} At leading
order, as for the case of the average multiplicity, only the first
term in eq.~(\ref{eq:derV1}) contributes. (The second term, having at
least one loop, starts at order $g^0$.) Moreover, the two Green's
functions $\Gamma^{(\pm)}$ should be evaluated at tree level, and the
wave-function renormalization factor $Z$ is equal to one.  The
tree-level 1-point functions
\begin{equation}
\Phi_\pm(z|x)\equiv\left.
\frac{\delta i{\cal V}_{_{SK}}[z|j_+,j_-]}{\delta j_\pm(x)}
\right|_{{j_+=j_-=j}\atop{\rm tree}}\; ,
\end{equation}
are represented diagrammatically by tree diagrams with one
external leg: 
\setbox1=\hbox to
1.55cm{\hfil\resizebox*{1.55cm}{!}{\includegraphics{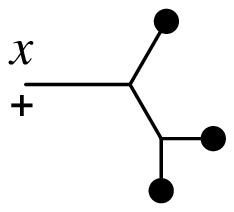}}}
\setbox2=\hbox to
1.55cm{\hfil\resizebox*{1.55cm}{!}{\includegraphics{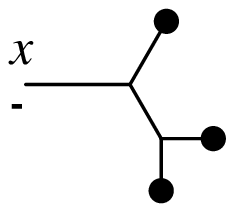}}}
\begin{equation}
\Phi_+(z|x)=\sum_{+/-}\;\raise -8mm\box1\;\;, \qquad
\Phi_-(z|x)= \sum_{+/-}\;\raise -8mm\box2
\quad.
\label{eq:Phi-pm}
\end{equation}
These graphs are evaluated with Schwinger-Keldysh rules wherein the 
off-dia\-go\-nal propagators are multiplied by a factor $z$ (see
eq.~(\ref{eq:modif-SK})).

In terms of these objects, eq.~(\ref{eq:deriv-gen-func}) can be written as 
\begin{eqnarray}
\left.\frac{{\cal F}^\prime(z)}{{\cal F}(z)}
\right|_{_{LO}}
&=&
\int d^4x d^4y\;G^0_{+-}(x,y)\,
(\square_x+m^2)(\square_y+m^2)\,\Phi_+(z|x)\Phi_-(z|y)
\nonumber\\
&=&
\int \frac{d^3\p}{(2\pi)^3 2E_\p}\;
\Big[
\int d^4x\; e^{ip\cdot x}\;(\square_x+m^2)\,\Phi_+(z|x)
\Big]
\nonumber\\
&&\qquad\qquad\qquad\times
\Big[
\int d^4y\; e^{-ip\cdot y}\;(\square_y+m^2)\,\Phi_-(z|y)
\Big]\; .
\end{eqnarray}
Finally, performing an integration by parts (as in eq.~(79) of
\cite{GelisV2}), we obtain  
\begin{eqnarray}
\left.
\frac{{\cal F}^\prime(z)}{{\cal F}(z)}
\right|_{_{LO}}
\!\!\!&=&\!\!\!
\int \frac{d^3\p}{(2\pi)^3 2E_\p}\;
\Big[
\int d^3\x\; e^{ip\cdot x}\;(\partial_x^0-iE_p)\,\Phi_+(z|x)
\Big]_{x_0=-\infty}^{x_0=+\infty}
\nonumber\\
&&\qquad\qquad\qquad\times
\Big[
\int d^3\y\; e^{-ip\cdot y}\;(\partial_y^0+iE_p)\,\Phi_-(z|y)
\Big]_{y_0=-\infty}^{y_0=+\infty}\, .
\nonumber\\
&&
\label{eq:der-V-5}
\end{eqnarray}
Therefore, at tree level, it is sufficient to find the behavior at
large times of the quantities $\Phi_\pm(z|x)$ to determine ${{\cal
F}^\prime(z)}/{{\cal F}(z)}$. It is important to note here that we
have to keep the contributions from both large positive and negative
times in eq.~(\ref{eq:der-V-5}). {\it A priori}, the fields
$\Phi_\pm(z|x)$ are not zero in the two limits.

Of course, the true difficulty lies in summing all the tree diagrams
that contribute to eqs.~(\ref{eq:Phi-pm}). A possible strategy for
performing this resummation is to reformulate the problem of summing
these diagrams as integral equations. In this case, $\Phi_\pm(z|x)$
can be expressed as\footnote{For $z=1$, we have
$\Phi_+(1|x)=\Phi_-(1|x)\equiv \phi_c(x)$; further, $G_{++}^0 -
G_{+-}^0 = G_{-+}^0 - G_{--}^0 = G_{_R}^0$, where $G_{_R}^0$ is the free
retarded Green's function. The two equations in this case are
equivalent and identically represent the solution of the classical
equation of motion.}
\begin{eqnarray}
\Phi_+(z|x)&=&i\int d^4y\; 
\Big\{G^0_{++}(x,y)\left[j(y)-\frac{g}{2}\Phi_+^2(z|y)\right]\nonumber\\
&&\qquad\qquad\qquad
-z\,G^0_{+-}(x,y)\left[j(y)-\frac{g}{2}\Phi_-^2(z|y)\right]
\Big\}
\nonumber\\
\Phi_-(z|x)&=&i\int d^4y\; 
\Big\{z\,G^0_{-+}(x,y)\left[j(y)-\frac{g}{2}\Phi_+^2(z|y)\right]\nonumber\\
&&\qquad\qquad\qquad
-G^0_{--}(x,y)\left[j(y)-\frac{g}{2}\Phi_-^2(z|y)\right]
\Big\}\; .
\label{eq:Phi-integral-eq}
\end{eqnarray}
Note that in these equations, the off-diagonal propagators $G^0_{+-}$
and $G^0_{-+}$ are multiplied by $z$. In principle, one could
solve these equations iteratively, starting from the $g=0$
approximation. Note also that $\Phi_\pm(z|x)$ obey the classical
equation of motion,
\begin{eqnarray}
(\square+m^2)\Phi_\pm(z|x)+\frac{g}{2}\Phi_\pm^2(z|x)=j(x)\; .
\label{eq:EOM-Phi}
\end{eqnarray}
This is easily seen from eqs.~(\ref{eq:Phi-integral-eq}) and from
the identities
\begin{eqnarray}
&&
(\square_x+m^2)\,G_{++}^0(x,y)=-i\delta(x-y)\; ,
\nonumber\\
&&
(\square_x+m^2)\,G_{--}^0(x,y)=+i\delta(x-y)\; ,
\nonumber\\
&&
(\square_x+m^2)\,G_{-+}^0(x,y)=(\square_x+m^2)\,G_{+-}^0(x,y)=0\; .
\label{eq:Green-identities}
\end{eqnarray}

\subsection{Calculation of $\Phi_\pm$ from the classical EOM}
\label{sec:EOM}
In the previous section, we expressed at leading order (tree level)
${{\cal F}^\prime(z)}/{{\cal F}(z)}$ in terms of a pair of fields,
$\Phi_\pm(z|x)$. The latter are specified by the integral equations
(\ref{eq:Phi-integral-eq}). We also pointed out that they are solutions of
the classical equation of motion. However, for this latter property to
be of any use in computing $\Phi_\pm(z|x)$, we must find the boundary
conditions they obey.

These can be determined by a method reminiscent of the standard
proof of Green's theorem. Multiply the 
equation of motion for $\Phi_+(z|x)$
\begin{equation}
(\stackrel{\rightarrow}{\square}_y+m^2)\Phi_+(z|y)
=j(y)-\frac{g}{2}\Phi_+^2(z|y)\; ,
\end{equation}
 by $G_{++}^0(x,y)$ on the left and the equation 
\begin{equation}
G^0_{++}(x,y)(\stackrel{\leftarrow}{\square}_y+m^2)=-i\delta(x-y)\; ,
\end{equation}
by $\Phi_+(z|y)$ on the right.  Integrate both the resulting
expressions over $y$, and subtract them to obtain
\begin{eqnarray}
\Phi_+(z|x)&=&i\int d^4y\;G^0_{++}(x,y)\;\Big[
j(y)-\frac{g}{2}\Phi_+^2(z|y)
\Big]
\nonumber\\
&&\qquad+i
\int d^4y\;G^0_{++}(x,y)\big[
\stackrel{\leftarrow}{\square}_y
  -\stackrel{\rightarrow}{\square}_y\big]\Phi_+(z|y)\; .
\label{eq:Green-1}
\end{eqnarray}

Following a similar procedure for the equation of motion of
$\Phi_-(z|x)$ and the expression in eq.~(\ref{eq:Green-identities})
for $G^0_{+-}(x,y)$, we obtain
\begin{eqnarray}
0&=&i\int d^4y\;G^0_{+-}(x,y)\;\Big[
j(y)-\frac{g}{2}\Phi_-^2(z|y)
\Big]
\nonumber\\
&&\qquad+i
\int d^4y\;G^0_{+-}(x,y)\big[
\stackrel{\leftarrow}{\square}_y
  -\stackrel{\rightarrow}{\square}_y\big]\Phi_-(z|y)
\end{eqnarray}
Multiplying this equation by $z$ and subtracting it from
eq.~(\ref{eq:Green-1}), we obtain
\begin{eqnarray}
\Phi_+(z|x)\!\!&=&
\!\!i\!\!\int\!\! d^4y\left\{G^0_{++}(x,y)\;\Big[
j(y)-\frac{g}{2}\Phi_+^2(z|y)
\Big]\right.\nonumber\\
&&\qquad\qquad
\left.-zG^0_{+-}(x,y)\;\Big[
j(y)-\frac{g}{2}\Phi_-^2(z|y)
\Big]\right\}
\nonumber\\
&&\qquad\qquad+i
\int d^4y\;G^0_{++}(x,y)\big[
\stackrel{\leftarrow}{\square}_y
  -\stackrel{\rightarrow}{\square}_y\big]\Phi_+(z|y)
\nonumber\\
&&\qquad\qquad-i
\int d^4y\;zG^0_{+-}(x,y)\big[
\stackrel{\leftarrow}{\square}_y
  -\stackrel{\rightarrow}{\square}_y\big]\Phi_-(z|y)\; .
\label{eq:Green-2}
\end{eqnarray}
The first two lines of this equation correspond to the integral
equation (\ref{eq:Phi-integral-eq}). This means that the sum of the
third and fourth lines must vanish.

We will now show that these terms indeed provide the boundary
conditions we are looking for.  They can be written as
\begin{eqnarray}
&&\!\!\!
0=\!\!\int\!\! d^4y\left\{
G^0_{++}(x,y)\big[
\stackrel{\leftarrow}{\square}_y
 \! -\!\stackrel{\rightarrow}{\square}_y\big]\Phi_+(z|y)
\!-\!zG^0_{+-}(x,y)\big[
\stackrel{\leftarrow}{\square}_y
  \!-\!\stackrel{\rightarrow}{\square}_y\big]\Phi_-(z|y)\right\}
\nonumber\\
&&
\!\!\!\!\!\!\!\!\!\!=\!\!
\int \!d^4y\,\partial_\mu^y\left\{
G^0_{++}(x,y)\big[
\stackrel{\leftarrow}{\partial}\!{}^\mu_y
  -\stackrel{\rightarrow}{\partial}\!{}^\mu_y\big]\Phi_+(z|y)
-zG^0_{+-}(x,y)\big[
\stackrel{\leftarrow}{\partial}\!{}^\mu_y
  -\stackrel{\rightarrow}{\partial}\!{}^\mu_y\big]\Phi_-(z|y)\right\}\, .
\nonumber\\
&&
\end{eqnarray}
These terms can be written as an integral over the boundary of space-time.
Assuming that the fields vanish at infinite distance in the
spatial directions\footnote{Alternatively, as is often the case in
numerical computations, one may assume periodic boundary conditions in
space corresponding to the spatial  topology of a
torus. The boundary integral in the spatial
directions can be dropped because a torus has no boundary.}, we are
left only with the integration at infinite times, with the 
temporal boundary conditions for $\Phi_\pm(z|x)$ at  $y_0=\pm\infty$ given by  
\begin{eqnarray}
&&
\int \!d^3\y\,\left\{
G^0_{++}(x,y)\big[
\stackrel{\leftarrow}{\partial}\!{}^0_y
  -\stackrel{\rightarrow}{\partial}\!{}^0_y\big]\Phi_+(z|y)
\right.
\nonumber\\
&&\qquad\qquad\left.
-zG^0_{+-}(x,y)\big[
\stackrel{\leftarrow}{\partial}\!{}^0_y
  -\stackrel{\rightarrow}{\partial}\!{}^0_y\big]
\Phi_-(z|y)\right\}_{y^0=-\infty}^{y^0=+\infty}
=0\; .
\label{eq:boundary1}
\end{eqnarray}
Note that this relation must be satisfied for any $x$. 

Proceeding in exactly the same way\footnote{The most notable difference is that
the equation obeyed by the propagator $G^0_{--}$ has the opposite sign
on its right hand side~:
\begin{equation*}
G^0_{--}(x,y)(\stackrel{\leftarrow}{\square}_y+m^2)=+i\delta(x-y)\; .
\end{equation*}
}
for $\Phi_-(z|x)$, one obtains 
\begin{eqnarray}
&&
\int \!d^3\y\,\left\{
zG^0_{-+}(x,y)\big[
\stackrel{\leftarrow}{\partial}\!{}^0_y
  -\stackrel{\rightarrow}{\partial}\!{}^0_y\big]\Phi_+(z|y)
\right.
\nonumber\\
&&\left.\qquad\qquad
-G^0_{--}(x,y)\big[
\stackrel{\leftarrow}{\partial}\!{}^0_y
  -\stackrel{\rightarrow}{\partial}\!{}^0_y\big]
\Phi_-(z|y)\right\}_{y^0=-\infty}^{y^0=+\infty}
=0\; .
\label{eq:boundary2}
\end{eqnarray}

The boundary conditions in eqs.~(\ref{eq:boundary1}) and (\ref{eq:boundary2})
are greatly simplified by expressing 
$\Phi_\pm(z|x)$ as a sum of plane waves, 
\begin{eqnarray}
&&
\Phi_+(z|x)\equiv\int\frac{d^3\p}{(2\pi)^3 2E_\p}
\left\{
f_+^{(+)}(z|x^0,\p)e^{-ip\cdot x}
+
f_+^{(-)}(z|x^0,\p)e^{ip\cdot x}
\right\}\; ,
\nonumber\\
&&
\Phi_-(z|x)\equiv\int\frac{d^3\p}{(2\pi)^3 2E_\p}
\left\{
f_-^{(+)}(z|x^0,\p)e^{-ip\cdot x}
+
f_-^{(-)}(z|x^0,\p)e^{ip\cdot x}
\right\}\; .\nonumber\\
&&
\label{eq:Phi-Fourier}
\end{eqnarray}
In these formulae, the variable $p_0$ is positive and equal to its
on-shell value $E_\p\equiv \sqrt{\p^2+m^2}$.  Because 
$\Phi_\pm(z|x)$ does not obey the free Klein-Gordon equation, the
coefficients functions must themselves depend on time. However, by
assuming that both the source $j(x)$ and the coupling constant $g$ are
switched off adiabatically at large negative and positive times, the
coefficient functions $f_\pm^{(\pm)}(z|x^0,\p)$ become constants in
the limit of infinite time. 

Substituting eq.~(\ref{eq:Phi-Fourier}) in eqs.~(\ref{eq:boundary1})
and (\ref{eq:boundary2}) and employing the explicit form of the
propagators $G^0_{\epsilon\epsilon^\prime}(x,y)$ in
eqs.~(\ref{eq:props}), these boundary conditions can be expressed much
more simply as boundary conditions of the $f_\pm^{(\pm)}(z|x^0,\p)$~:
\begin{eqnarray}
&&
f_+^{(+)}(z|x^0=-\infty,\p)=0\; ,
\nonumber\\
&&
f_-^{(-)}(z|x^0=-\infty,\p)=0\; ,
\nonumber\\
&&
f_-^{(+)}(z|x^0=+\infty,\p)=
z\,f_+^{(+)}(z|x^0=+\infty,\p)\; ,
\nonumber\\
&&
f_+^{(-)}(z|x^0=+\infty,\p)=
z\,f_-^{(-)}(z|x^0=+\infty,\p)\; .
\label{eq:boundary3}
\end{eqnarray}
These relations must be satisfied for each momentum mode $\p$. 

Eq.~(\ref{eq:der-V-5}) can be
rewritten in terms of the coefficient functions
$f_\pm^{(\pm)}(z|x_0,\p)$ at $x_0=+\infty$ as 
\begin{equation}
\left.
\frac{{\cal F}^\prime(z)}{{\cal F}(z)}
\right|_{_{LO}}
=
\int \frac{d^3\p}{(2\pi)^3 2E_\p}\;
f_+^{(+)}(z|+\infty,\p)\,f_-^{(-)}(z|+\infty,\p)\; .
\label{eq:dF1}
\end{equation}
This equation, the boundary conditions in 
eqs.~(\ref{eq:boundary3}) and the fact that the fields $\Phi_\pm$ obey
the classical equation of motion, uniquely determine the
generating function at leading order.

\subsection{Evaluating eq.~(\ref{eq:dF1}): practical considerations}
We showed in the previous subsection that the eqs.~(\ref{eq:EOM-Phi}),
(\ref{eq:Phi-Fourier}), (\ref{eq:dF1}) and the boundary conditions in
eqs.~(\ref{eq:boundary3}) completely map the problem of finding the
generating function at leading order into the problem of finding
certain solutions of the classical equations of motion. In a sense,
this is the analog of what was achieved in \cite{GelisV2} for the
average number of produced particles at leading order. The only
differences are that the boundary conditions now depend on the
argument $z$ of the generating function and that the $\Phi_\pm$ are
required to satisfy boundary conditions at $x_0=+\infty$.

Indeed, for $z=1$, ${\cal F}^\prime(1)/{\cal F}(1)$ is nothing but
$\big<n\big>$. We should therefore recover the results of
\cite{GelisV2}, namely, that $\Phi_+(1|x)$ and $\Phi_-(1|x)$ are equal
and that they are both the {\bf retarded} solution of the classical
EOM with a vanishing initial condition. It is easy to see that this
result is implied by the eqs.~(\ref{eq:boundary3}). The last two
equations tell us that $\Phi_+(1|x)$ and $\Phi_-(1|x)$ have the same
coefficient functions at $x^0=+\infty$, implying that $\Phi_+(1|x)$
and $\Phi_-(1|x)$ are identical everywhere~:
\begin{equation}
\forall x\; ,\quad \Phi_+(1|x)=\Phi_-(1|x)\; .
\end{equation}
 Likewise, the first two boundary equations tell us that
\begin{equation}
\Phi_+(1|x^0=-\infty,\x)=\Phi_-(1|x^0=-\infty,\x)=0\; .
\end{equation}
Therefore $\Phi_\pm(1|x)$ is nothing but the retarded solution of
the equation of motion with a vanishing initial condition.

Although we found the boundary conditions for $\Phi_\pm(z|x)$, finding
the two solutions of the classical equation of motion that fulfill
these boundary conditions is much more complicated than finding the
retarded solution. The reason for the complication is that the
boundary conditions in this case are expressed partly at $x^0=-\infty$
and partly at $x^0=+\infty$. In practice, this means that the problem
cannot be solved as an initial value problem starting from some known
value at $x_0=-\infty$ and solving the EOM forward in time. Instead,
most of the methods for solving this kind of problem numerically are
``relaxation processes''. Generically, one perceives them as
algorithms where one introduces some fictitious ``relaxation time''
variable $\xi$. The simulation begins at $\xi=0$ with functions
$\Phi_\pm$ that satisfy all the boundary conditions (it is easy to
construct such fields) but not the equation of motion. These fields
are then evolved in $\xi$ according to the equation
\begin{equation}
\partial_\xi \Phi_\pm = (\square_x+m^2)\Phi_\pm +
\frac{g}{2}\Phi_\pm^2-j(x)\; ,
\end{equation}
which admits solutions of the EOM as fixed points. The right hand side
of the previous equation can in principle be replaced by any function
that vanishes when $\Phi_\pm$ is a solution of the classical EOM. The
freedom to chose this function could be used in order to ensure that
the fixed point is attractive.  Finally, at each step in the
fictitious time $\xi$, we only need to make sure that the updating
procedure preserves the boundary conditions. A numerical algorithm
that implements this procedure will be discussed in future. We note
that somewhat similar techniques have been developed recently to study
the non-equilibrium real time properties of quantum
fields~\cite{BergeS3}.

\section{Variance at Leading Order}
\label{sec:variance}
Although it is numerically challenging to find a pair of solutions of
the classical EOM that obeys the boundary conditions given in
eqs.~(\ref{eq:boundary3}), the results of the previous section are
nevertheless very useful in deriving an expression for the variance of the 
multiplicity distribution, defined as 
\begin{equation}
\sigma \equiv \big<n^2\big>-\big<n\big>^2\; .
\label{eq:var-def}
\end{equation}
The starting point is to differentiate eq.~(\ref{eq:dF1}) {\sl with
respect to $z$}, and evaluate the result at $z=1$. This gives the
following relation
\begin{eqnarray}
&&
\left.\sigma-\big<n\big>\right|_{_{LO}}=
\left.
{\cal F}^{\prime\prime}(1)-\Big({\cal F}^\prime(1)\Big)^2
\right|_{_{LO}}
\nonumber\\
&&
\qquad\qquad
=
\int \frac{d^3\p}{(2\pi)^3 2E_\p}\;
\Big[
f^{(-)}(1|+\infty,\p)\,f_+^{(+)\prime}(1|+\infty,\p)
\nonumber\\
&&\qquad\qquad\qquad\qquad\qquad
\smash{+
f^{(+)}(1|+\infty,\p)\,f_-^{(-)\prime}(1|+\infty,\p)
}\Big]\; .
\label{eq:dF2}
\end{eqnarray}
(We used the identity ${\cal F}(1)=1$ to get
rid of the denominators.) The prime symbol $\prime$ denotes the
differentiation with respect to $z$. Moreover, we simplified the
formula by taking into account the fact that, at $z=1$, the two fields
$\Phi_\pm$ are equal and thus have the same Fourier
coefficients; we did not therefore spell out their
subscripts $\pm$. (Note however that the $z$-derivatives of these Fourier coefficients may differ.)

The next step is to recognize that the derivatives (with respect to $z$)--
$f_\pm^{(\pm)\prime}$ -- of the Fourier coefficients of $\Phi_\pm$ are
the Fourier coefficients of the derivatives $\Phi_\pm^\prime$. 
Differentiating the EOM in eq.~(\ref{eq:EOM-Phi}) with respect to $z$,
one readily sees that these derivatives obey the EOM for small
fluctuations on top of the classical field, namely, 
\begin{eqnarray}
\Big[\square+m^2+g\Phi(x)\Big]\Phi_\pm^\prime(1|x)=0\; ,
\label{eq:EOM-eta}
\end{eqnarray}
where we denote by $\Phi(x)\equiv \Phi_+(1|x)=\Phi_-(1|x)$,  the retarded solution of the classical EOM with
null initial conditions. Moreover, one can also differentiate with
respect to $z$ the boundary conditions of eqs.~(\ref{eq:boundary3}) to find in turn the boundary conditions satisfied by the fields
$\Phi_\pm^\prime(1|x)$. We obtain
\begin{eqnarray}
&&
f_+^{(+)\prime}(1|x^0=-\infty,\p)=0\; ,
\nonumber\\
&&
f_-^{(-)\prime}(1|x^0=-\infty,\p)=0\; ,
\nonumber\\
&&
f_-^{(+)\prime}(1|x^0=+\infty,\p)=
f_+^{(+)\prime}(1|x^0=+\infty,\p)+f^{(+)}(1|x^0=+\infty,\p)\; ,
\nonumber\\
&&
f_+^{(-)\prime}(1|x^0=+\infty,\p)=
f_-^{(-)\prime}(1|x^0=+\infty,\p)+f^{(-)}(1|x^0=+\infty,\p)\; .
\nonumber\\
&&
\label{eq:boundary4}
\end{eqnarray}
Although we still have boundary conditions both at $x_0=-\infty$ and
at $x_0=+\infty$, the problem can be solved as an initial value
problem, in contrast, as we have seen, with the case of the generating
function. This is because the equation of motion for small field
fluctuations, eq.~(\ref{eq:EOM-eta}), is {\bf linear}. The fields
$\Phi_\pm^\prime(1|x)$ that obey these boundary conditions can
therefore be determined by introducing a linear basis for the small
fluctuation fields. To do so, first introduce two fields
$\eta_{\pm\vec\q}(x)$ obeying the EOM for small fluctuations, that are
equal to plane waves when $x_0\to-\infty$, namely,
\begin{eqnarray}
&&
\Big[\square_x+m^2+g\Phi(x)\Big]\eta_{\pm\q}(x)=0\; ,
\nonumber\\
&&
\eta_{\pm\q}(x)=e^{\pm i q\cdot x}\qquad\mbox{when}\quad x_0\to -\infty\; .
\end{eqnarray}
This initial condition for $\eta_{\pm\q}(x)$ is permitted provided the
momentum $q$ is on-shell because the classical field $\Phi(x)$
vanishes in the remote past. From the first two of
eqs.~(\ref{eq:boundary4}), we know that $\Phi_+^\prime(1|x)$ has no
positive energy component and $\Phi_-^\prime(1|x)$ has no negative
energy component at $x_0=-\infty$. Therefore $\Phi_+^\prime(1|x)$ must
be a linear combination of the $\eta_{+\q}$'s while
$\Phi_-^\prime(1|x)$ is a linear combination of the $\eta_{-\q}$'s~:
\begin{eqnarray}
&&
\Phi_+^\prime(1|x)=\int \frac{d^3\q}{(2\pi)^3 2E_\q} \;
C_{+\q}\;\eta_{+\q}(x)
\; ,
\nonumber\\
&&
\Phi_-^\prime(1|x)=\int \frac{d^3\q}{(2\pi)^3 2E_\q} \;
C_{-\q}\;\eta_{-\q}(x)
\; .
\end{eqnarray}
The coefficients $C_{\pm\q}$ can be determined from the boundary
conditions at $x_0=+\infty$ by decomposing the fields
$\eta_{\pm\q}(x)$ into positive and negative energy plane
waves\footnote{Although of course, at $x_0=-\infty$, they contain
  only one of the two components, they acquire the other component
  during the evolution through the background classical field.}. With
the Fourier decomposition
\begin{eqnarray}
&&
\eta_{+\q}(x)\equiv
\int \frac{d^3\p}{(2\pi)^3 2E_\p}
\Big\{
h_{+\q}^{(+)}(x_0,\p)\,e^{-ip\cdot x}
+
h_{+\q}^{(-)}(x_0,\p)\,e^{ip\cdot x}
\Big\}\; ,
\nonumber\\
&&
\eta_{-\q}(x)\equiv
\int \frac{d^3\p}{(2\pi)^3 2E_\p}
\Big\{
h_{-\q}^{(+)}(x_0,\p)\,e^{-ip\cdot x}
+
h_{-\q}^{(-)}(x_0,\p)\,e^{ip\cdot x}
\Big\}\; ,
\end{eqnarray}
we may now  rewrite the boundary conditions at $x_0=+\infty$ in
terms of the Fourier modes. We obtain,
\begin{eqnarray}
&&
\int\!\!\frac{d^3\q}{(2\pi)^32E_\q}
\Big[h_{-\q}^{(+)}(+\infty,\p)\,C_{-\q}-h_{+\q}^{(+)}(+\infty,\p)\,C_{+\q}\Big]
=f^{(+)}(1|+\infty,\p)
\; ,
\nonumber\\
&&
\int\!\!\frac{d^3\q}{(2\pi)^32E_\q}
\Big[h_{+\q}^{(-)}(+\infty,\p)\,C_{+\q}-h_{-\q}^{(-)}(+\infty,\p)\,C_{-\q}\Big]
=f^{(-)}(1|+\infty,\p)
\; .
\nonumber\\
&&
\label{eq:boundary5}
\end{eqnarray}
These equations must be satisfied for all the momenta $\p$ and are a
linear system of equations to determine the coefficients $C_{\pm\q}$.
Once they have been found, one can express the variance at leading
order in terms of these previously introduced objects
\begin{eqnarray}
&&
\left.\sigma-\big<n\big>\right|_{_{LO}}
=
\int\frac{d^3\p}{(2\pi)^32E_\p}\frac{d^3\q}{(2\pi)^32E_\q}\;
\Big\{
f^{(-)}(1|+\infty,\p)\,h_{+\q}^{(+)}(+\infty,\p)\,C_{+\q}
\nonumber\\
&&\qquad\qquad\qquad\qquad\qquad\qquad
+
f^{(+)}(1|+\infty,\p)\,h_{-\q}^{(-)}(+\infty,\p)\,C_{-\q}
\Big\}\; .
\label{eq:variance-final}
\end{eqnarray}

In conclusion, the variance (at leading order) of the number of produced
particles, may be computed as follows:
\begin{itemize}
\item[(i)] Find the solution $\Phi(x)$ of the classical equation of
motion, with null initial conditions, and calculate its Fourier modes
$f(\pm)(+\infty,\p)$ at large positive times. This computation will in
principle have already been done when evaluating numerically the 
average multiplicity at leading order. 

\item[(ii)] Find the functions $\eta_{\pm\q}(x)$ by solving the EOM
(\ref{eq:EOM-eta}) for small field fluctuations about the classical
field $\Phi(x)$, with initial conditions $\exp(\pm iq\cdot
x)$. Usually, one will have discretized the spatial volume, and this
step must be carried out for each momentum $\q$ of the dual lattice.

\item[(iii)] The next step (in analogy with the procedure in (i)) is
  to calculate the Fourier modes $h_{\pm\q}^{(\pm)}(+\infty,\p)$ of
  the $\eta_{\pm\q}$' at large positive times.

\item[(iv)] With $f(\pm)(+\infty,\p)$ and
  $h_{\pm\q}^{(\pm)}(+\infty,\p)$ in hand, solve the linear system of
  equations (\ref{eq:boundary5}) by inverting a matrix (albeit a large
  one) to find the coefficients $C_{\pm\q}$.

\item[(v)] Finally, one obtains the variance from
eq.~(\ref{eq:variance-final}).

\end{itemize}
As one can see, the calculation of the variance requires a lot more
computational work than the calculation of the average multiplicity.
However, unlike in the case of the generating function itself, all the
partial differential equations that need to be solved are with
retarded boundary conditions. All the other steps are
``elementary'' because they involve only Fourier decompositions or the 
matrix inversion of a linear system of equations.

\section{Generating function\\ for the energy distribution}
\label{sec:E-GF}
Thus far, we only discussed generating functions for 
the probabilities for producing a given number of particles.
However, the distribution of radiated
energy\footnote{Although we focus here on the generating
function for the distribution of energy, our derivation holds 
for any other quantity that is additive for a set of $n$
particles.} may be better defined in the
infrared for gauge theories with massless particles.

We denote $P(E)$ as the probability density for the energy with $P(E)dE$ being 
the probability of radiating energy between the
values $E$ and $E+dE$. One may write $P(E)$ in terms
of transition amplitudes as 
\begin{equation}
P(E)=
\sum_{n=0}^\infty
\frac{1}{n!}
\int \prod_{i=1}^n \frac{d^3\p_i}{(2\pi)^3 2E_i}
\;\delta(E-\sum_{i=1}^n E_i)
\left|
\big<\p_1\cdots \p_n\,{}_{\rm out}\big|0_{\rm in}\big>
\right|^2\; .
\label{eq:PA}
\end{equation}
In this formula, $E_i$ is the energy of the particle of momentum
$\p_i$. 

Replacing  the delta function
in eq.~(\ref{eq:PA}) by the identity 
\begin{equation}
\delta(E-\sum_{i=0}^n E_i)
=\frac{1}{2\pi}\int\limits_{-\infty}^{+\infty}
d\theta \; e^{i\theta\left(\sum_{i} E_i-E\right)}\; ,
\end{equation}
we obtain 
\begin{equation}
P(E)=\frac{1}{2\pi}\int\limits_{-\infty}^{+\infty}
d\theta \; e^{-i\theta E}\;{\cal F}_{_E}(\theta)\; ,
\end{equation}
where the generating function ${\cal F}_{_E}(\theta)$ is defined as
\begin{equation}
{\cal F}_{_E}(\theta)
\equiv
\sum_{n=0}^\infty
\frac{1}{n!}
\int \prod_{i=1}^n \frac{d^3\p_i}{(2\pi)^3 2E_i}
e^{i\theta E_i}
\;
\left|
\big<\p_1\cdots \p_n\,{}_{\rm out}\big|0_{\rm in}\big>
\right|^2\; .
\label{eq:FA}
\end{equation}
It is straightforward to obtain for this generating function a form
similar to eq.~(\ref{eq:genF-1})~:
\begin{equation}
{\cal F}_{_E}(\theta)=
\left.
e^{{\cal D}^{E}_\theta[j_+,j_-]}\;e^{iV[j_+]}\; e^{-iV^*[j_-]}
\right|_{j_+=j_-=j}\; ,
\label{eq:genFA}
\end{equation}
where we denote
\begin{equation}
{\cal D}^{E}_\theta[j_+,j_-]\equiv
\frac{1}{Z}\int d^4x d^4y\; G_{+-,\theta}^{0,E}(x,y)\;
(\square_x+m^2)(\square_y+m^2)\;
\frac{\delta}{\delta j_+(x)}\,\frac{\delta}{\delta j_-(y)}\; ,
\end{equation}
with
\begin{equation}
G_{+-,\theta}^{0,E}(x,y)
\equiv
\int \frac{d^3\p}{(2\pi)^3 2E_\p}\;e^{i\theta E_\p}\;e^{ip\cdot(x-y)}\; .
\end{equation}
This means that the generating function ${\cal F}_{_E}(\theta)$ is the
sum of the vacuum-vacuum diagrams in a Schwinger-Keldysh formalism in
which the off-diagonal propagators have been modified as follows~:
\begin{eqnarray}
&&
G_{+-}^0\quad\longrightarrow\quad G_{+-,\theta}^{0,E}\; ,
\nonumber\\
&&
G_{-+}^0\quad\longrightarrow\quad G_{-+,\theta}^{0,E}\; .
\label{eq:subs3}
\end{eqnarray}
At leading order, the derivative of $\ln{\cal F}_{_E}(\theta)$ can be
written in terms of a pair of solutions of the classical equation of
motion, with the following boundary conditions~:
\begin{eqnarray}
&&
f_+^{(+)}(\theta|x^0=-\infty,\p)=0\; ,
\nonumber\\
&&
f_-^{(-)}(\theta|x^0=-\infty,\p)=0\; ,
\nonumber\\
&&
f_-^{(+)}(\theta|x^0=+\infty,\p)=
e^{i\theta E_\p}\,f_+^{(+)}(\theta|x^0=+\infty,\p)\; ,
\nonumber\\
&&
f_+^{(-)}(\theta|x^0=+\infty,\p)=
e^{i\theta E_\p}\,f_-^{(-)}(\theta|x^0=+\infty,\p)\; .
\label{eq:boundary6}
\end{eqnarray}
As one can see, calculating the generating function for the energy
distribution involves solving the classical equation of motion with 
boundary conditions similar to those for the number distributions.

\section{Conclusions}
In this paper, we discussed the generating function for the
distribution of produced particles in a field theory coupled to a
strong time-dependent source. We obtained a general formula for the
logarithmic derivative of the generating function. At leading order,
this formula can be expressed in terms of a pair of solutions of the
classical EOM. We found that these solutions must obey boundary
conditions both at the initial and final time. Finding numerical
solutions of a non-linear EOM that obey these boundary conditions is
much more difficult that solving the EOM with retarded boundary
conditions.  At present, this problem is unsolved; we speculate that
numerical ``relaxation methods" may be applied to solve this
problem. 

From the results obtained for the generating function, we also
sketched an algorithm for calculating the variance of the number of
produced particles, which only involves partial differential equations
with retarded boundary conditions.

Finally, we showed that the generating function for the
distribution of produced energy may in principle be calculated by very
similar methods.

In addition to the numerical investigations alluded to in the previous
paragraphs, several extensions of this work are being considered.  One
of these is to generalize the present study to the situation where one
considers particle production only in some restricted portion of the
phase-space. When particle production is forbidden in the
complementary part of phase-space, the corresponding generating
function could be used in order to study the precise relation of
colorless partonic configurations in the color glass condensate
framework to experimentally observed rapidity gaps.  Another possible
extension of this work is to derive an evolution equation that would
drive the dependence of the generating function with the center of
mass energy of the collision under consideration.

\section*{Acknowledgements}
We would like to thank J.~Berges, D.~Dietrich, S.~Jeon, K.~Kajantie, T.~Lappi,
L.~McLerran and J.-Y.~Ollitrault for useful discussions on closely
related issues. FG and RV would like to thank the hospitality of the Physics
Department at McGill University, where part of this work was 
completed. RV's research is supported by DOE Contract No. DE-AC02-98CH10886.

\bibliographystyle{unsrt} 

\end{document}